\pdfoutput=1
\documentclass[twocolumn,prb,aps,longbibliography,showpacs,
superscriptaddress,10pt]{revtex4-1}

\usepackage{amsfonts,amssymb,dsfont}
\usepackage[sumlimits,intlimits]{amsmath}
\usepackage{graphics}
\usepackage{graphicx}
\usepackage{color}
\usepackage{mathrsfs}
\usepackage{textcomp}
\usepackage{verbatim}
\usepackage{hyperref}    
\usepackage{bm}          
\usepackage{braket}

\def\cH{\hat{\cal H}}

\def\cO{{\cal O}}

\def\bk{{\bf k}}

\def\bq{{\bf q}}
\def\bQ{{\bf Q}}
\def\bp{{\bf p}}

\def\br{{\bf r}}
\def\bR{{\bf R}}

\def\bzeta{\boldsymbol \zeta}

\def\hA{\hat A}
\def\ha{\hat a}

\def\hB{\hat B}

\def\hPsi{\hat \Psi}

\begin{document}

\title{Interaction-induced transition in the quantum chaotic dynamics of a disordered metal}

\author{S.V.~Syzranov}
\affiliation{Physics Department, University of California, Santa Cruz, CA 95064, USA}
\affiliation{Joint Quantum Institute, NIST/University of Maryland, College Park, MD 20742, USA}

\author{A.V.~Gorshkov}
\affiliation{Joint Quantum Institute, NIST/University of Maryland, College Park, MD 20742, USA}
\affiliation{Joint Center for Quantum Information and Computer Science, NIST/University of Maryland, College Park, MD 20742, USA} 

\author{V.M.~Galitski}
\affiliation{Joint Quantum Institute, NIST/University of Maryland, College Park, MD 20742, USA}
\affiliation{Condensed Matter Theory Center, Physics Department, University of Maryland, College Park, MD 20742, USA}

\date{\today}

\begin{abstract}
	We demonstrate that a weakly disordered metal with short-range interactions exhibits a transition
in the quantum chaotic dynamics when changing the temperature or the interaction strength.
For weak interactions,
the system displays exponential growth of the out-of-time-ordered correlator (OTOC)
of the current operator.
The Lyapunov exponent of this growth is temperature-independent in the limit of vanishing interaction.
With increasing the temperature or the interaction strength, the system undergoes a transition to a non-chaotic
behaviour, for which the exponential growth of the OTOC is absent.
We conjecture that the transition manifests itself in the 
quasiparticle energy-level statistics and 
also discuss ways of its explicit observation in cold-atom setups. 
\end{abstract}

\maketitle 

A classical chaotic system
is a system whose evolution is exponentially sensitive to the initial conditions. 
Because small perturbations of the Hamiltonian or the initial state 
of a chaotic system may dramatically change its dynamics, its evolution may appear random
even without any random elements in the Hamiltonian.

The concept of chaos in {\it quantum} systems is more subtle and has several different
widely used definitions.
When a quantum system has a well-defined classical limit, this system is often called chaotic
if the classical limit of its dynamics is chaotic.
Another definition of quantum chaos involves energy-level statistics. When the dynamics of 
a system is apparently random, one may expect this system to be described by the random-matrix theory,
which leads to the Wigner-Dyson statistics of the energy levels.
An immense amount of numerical data (see Ref.~\onlinecite{Efetov:book} for review)
suggests that various systems with Wigner-Dyson level statistics,
such as disordered metals and non-integrable billiards, exhibit classical chaotic dynamics, thereby
confirming the equivalence of the two definitions in these cases.

Numerous recent studies suggest another notion of quantum chaos, which is related to the exponential growth
of out-of-time-ordered correlators (OTOCs) of Hermitian operators,
quantities of the form $\langle [\hA(t),\hB(0)]^2\rangle$. Such correlators were first introduced
half a century ago~\cite{LarkinOvchnnikov} for electrons in weakly disordered metals; 
it was demonstrated that the correlator of the single-momentum
projections grows exponentially, $\langle[\hat p_z(t),\hat p_z(0)]^2\rangle\propto\exp(2\lambda t)$, where
the exponent $\lambda$ gives the rate of divergence between two initially close classical electron trajectories.
However, exponential growth of OTOCs in quantum systems takes place
only on sufficiently short times, in contrast with classical chaotic dynamics.

The last couple of years have seen an upsurge of research 
activity (see, for example, Refs.~\onlinecite{Maldacena:bound,AleinerFaoroIoffe,Fan:OTOCMBL,Chen:MBLscrambling,SwingleChowdhury:LocScrambling,HuangChen:OTOCMBL,Rosenbaum:rotor,PatelSachdev,BagretsKamenev,Kivelson:phonons,Yao:2copies,Knap:2copies,Swingle:measurement,ZhuHafeziGrover:measureClock,Danshita:SYKmeasurement,GarttnerRey:ionOTOC,Tsuji:measurement,Li:NMRmeas,YungerHalpern:kindaReview,YungerHalpern:Jarzinsky,Garcia:SYKtransition}) on OTOCs and quantum chaos, in part
motivated by a recent prediction~\cite{Maldacena:bound} 
of the bound $\lambda\leq2\pi T/\hbar$
on the Lyapunov exponent
in strongly interacting systems at temperature
$T$.
OTOCs are also expected~\cite{Fan:OTOCMBL,Chen:MBLscrambling,SwingleChowdhury:LocScrambling,HuangChen:OTOCMBL,PatelSachdev} 
to distinguish between many-body-localised~\cite{BAA}\footnote{See Ref.~\onlinecite{NandkishoreHuse:review} for a review} 
and many-body-delocalised phases. Despite recent advances in rigorous microscopic 
calculations of OTOCs in
 disordered interacting systems (see, e.g., Refs.~\onlinecite{AleinerFaoroIoffe,PatelSachdev,BagretsKamenev,Kivelson:phonons}),
the generic conditions for the existence of chaotic behaviour in such systems still remain to be investigated.

In this paper, we demonstrate that weakly disordered interacting systems exhibit a phase transition
from quantum-chaotic to non-chaotic behaviour when increasing the temperature or the interaction strength.
We distinguish between
chaotic and non-chaotic behaviour via the OTOC of momentum projections.

{\it Results.} We demonstrate that in a weakly disordered system with short-range interactions,
the time dependence of the OTOC of momentum $\hat P_z$ of the entire system is given by
\begin{align}
	F(t)=
	\left<[\hat P_z(t),\hat P_z(0)]^2\right>
	\propto \exp\left[{2\lambda t-\frac{2t}{\tau(T)}}\right],
	\label{Main}
\end{align}
where we omitted the subleading terms, which at $t=0$ ensure the vanishing of the correlator;
the averaging $\langle\ldots\rangle$ is carried out with respect to both the state of the electrons 
and disorder realisations;
$T$ is the temperature;
$\lambda$ is the classical Lyapunov exponent in a non-interacting disordered system;
and $1/\tau(T) \propto T^2$ is the rate of inelastic quasiparticle scattering at the Fermi surface.
Depending on which term in the exponent dominates, the OTOC exhibits exponential growth or lack thereof.

At low temperatures and interaction strengths, the behaviour of the system is similar 
to that in the absence of interactions, with an exponentially growing OTOC, 
which comes from pairs of close electron trajectories and from energies in the
interval $\varepsilon\sim T$ near the Fermi surface.
The Lyapunov exponent $\lambda$ is determined by the quasiparticle parameters at the Fermi surface 
and is temperature-independent, in contrast with strongly-interacting systems,
for which, under certain assumptions,
 a temperature-dependent 
bound on the Lyapunov exponent was predicted~\cite{Maldacena:bound}.
The exponential growth only persists for $t$ shorter than the Ehrenfest time 
$t_E=\lambda^{-1}\ln(a p_F)$ (hereinafter $\hbar=1$), where $a$ is the characteristic impurity size
and $p_F$ is the Fermi momentum.

For large temperatures
and strong interactions, inelastic large-momentum scattering destroys correlations
between electrons with close trajectories and thus suppresses
the exponential growth. There exists a critical temperature (or interaction strength, for a given temperature)
which corresponds to $\tau(T)=\lambda^{-1}$ and separates the regimes of exponential growth and the lack of exponential
growth.
We conjecture that the transition from chaotic to non-chaotic behaviour is accompanied by
a change of quasiparticle energy-level statistics from Wigner-Dyson to Poisson or to other types
of statistics. 
The transition may also be observed explicitly in double layers of ultracold atoms or molecules exposed to a random potential.

{\it Model.} We consider a weakly disordered metal in dimension $d>1$ described by the Hamiltonian 
\begin{align}
	\cH	=
	&\int\hPsi^\dagger(\br)
	\left[\xi_{\hat\bk}+\sum_i U(\br-\br_i)\right]\hPsi(\br)\:d\br
	\nonumber\\
	&+\int \hPsi^\dagger(\br)\hPsi^\dagger(\br^\prime)
	w(\br-\br^\prime)\hPsi(\br^\prime)\hPsi(\br)\,d\br\: d\br^\prime,
\end{align}
where $\hPsi^\dagger$ and $\hPsi$ are fermionic operators, $\xi_{\hat\bk}$ is the operator
of the kinetic energy; there are identical randomly located impurities in the system,
with $U(\br-\br_i)$ being the
potential of the $i$-th impurity; $w(\br-\br^\prime)$
is the interaction potential between two particles, which is assumed to be short-range in this paper.
Here we disregard the spin degree of freedom, because it has no qualitative effect on our results.

{\it Formalism.} Describing transport in a disordered system often involves
perturbative expansions in interactions and random potential, using Wick's theorem to reduce
observables in various orders of perturbation theory
to two-point correlators, i.e. Green's functions~\cite{AGD}.
A similar approach may be adopted when calculating four-point OTOCs~\cite{AleinerFaoroIoffe},
defining Green's functions on a four-branch time contour
instead of the conventional two-branch Keldysh contour~\cite{Kamenev:book}.

Here we use an alternative approach, developed recently in Ref.~\onlinecite{Syzranov:OTOCdot},
which consists in deriving kinetic (or master) equations for higher-order correlators,
similar to the joint distribution functions of two copies of the system of electrons. 
We introduce four correlation functions
\begin{align}
   &K^{\alpha\beta}(\br_1,\br_{1^\prime};\br_2,\br_{2^\prime};t)=
   \nonumber\\
   &\left<\left[\hPsi^{\alpha\dagger}(\br_1,t)\hPsi^\alpha(\br_{1^\prime},t),
   \hat P_z(0)\right]
   \left[\hPsi^{\beta\dagger}(\br_2,t)\hPsi^\beta(\br_{2^\prime},t),
   \hat P_z(0)\right]\right>,
   \label{KAlphaBeta}
\end{align}
where each of the indices $\alpha$ and $\beta$ may take two values: $e$ (electron)
or $h$ (hole); $\hPsi^e(\br)=\hPsi(\br)$ is the annihilation operator for an electron
at location $\br$ and $\hPsi^h=\hPsi^\dagger(\br)$ is the hole annihilation operator.
The time evolution of the correlators $K^{ee}$, $K^{hh}$ and $K^{eh,he}$ 
is similar to the evolution
of the joint distribution
functions of two electrons, two holes and electron-hole
pairs, respectively.
Moreover, for a system in a classical environment,
of which a random potential
is a special case, the evolution of the correlators (\ref{KAlphaBeta})
exactly matches that of the density matrix of pairs of electrons and/or holes, respectively~\cite{Syzranov:OTOCdot}.

The relation of the OTOC of momentum projections to the correlation functions $K^{\alpha\beta}$ is given by
\begin{align}
	F(t)=
	&\int_{\bR_1,\bR_2,\bp_1,\bp_2} p_{1z}p_{2z}\: K^{ee}(\bp_1,\bR_{1};\bp_2,\bR_{2};t),
	\label{FviaK}
\end{align}
where our conventions for coordinate and momentum integration are 
$\int_\bR\ldots=\int\ldots d\bR$ and $\int_\bp\ldots=\int\ldots\frac{d\bp}{(2\pi)^d}$, respectively,
in the $d$-dimensional space;
$K^{ee}(\bp_1,\bR_{1};\bp_2,\bR_{2};t)$ is the Wigner-transform (a function of the centre-of-mass
coordinates $\bR_i=\frac{\br_{i}+\br_{i^\prime}}{2}$ and the Fourier-transform
of the coordinate difference, $\br_i-\br_{i^\prime}\rightarrow \bp_i$) of the correlation
function $K^{ee}(\br_1,\br_{1^\prime};\br_2,\br_{2^\prime};t)$. Equation~(\ref{FviaK})
is similar to the relation between the correlator $\langle p_{1z}p_{2z}\rangle$
of the momentum projections of two electrons and the joint density matrix of these
electrons.

The initial values of the correlators (\ref{KAlphaBeta}) for an electron gas with
the (single-particle) distribution function $f_\bk$ are given by
\begin{align}
	K^{\alpha\alpha}(\bp_1,\bR_{1};\bp_2,\bR_{2};0)
	=-K^{\alpha\bar\alpha}(\bp_1,\bR_{1};\bp_2,\bR_{2};0)
	\nonumber\\
	=\partial_{Z_1}\partial_{Z_2}g(\bR_1-\bR_2,\bp_1-\bp_2),
	\label{gZZ}
\end{align}
where $\bar\alpha$ labels an ``antiparticle'' of $\alpha$ ($\bar e=h$; $\bar h=e$); 
$\partial_{Z_1}$ and $\partial_{Z_2}$ are derivatives with respect to
the $z$ components of $\bR_1$ and $\bR_2$; the function $g(\bR_1-\bR_2,\bp_1-\bp_2)$
is given by
\begin{align}
  g(\bR_1-\bR_2,\bp_1,\bp_2)=
  -(4\pi)^{d}\delta(\bp_1-\bp_2)
  \nonumber\\
  \int_{\bq}f_{\bp_1-\bq}(1-f_{\bp_1+\bq})
  e^{2i\bq\cdot(\bR_1-\bR_2)}
  \label{gDef}
\end{align}
and is sharply peaked at the origin as a function of $\bR_1-\bR_2$ and $\bp_1-\bp_2$.

\begin{figure}[t]
	\centering
	\includegraphics[width=0.32\textwidth]{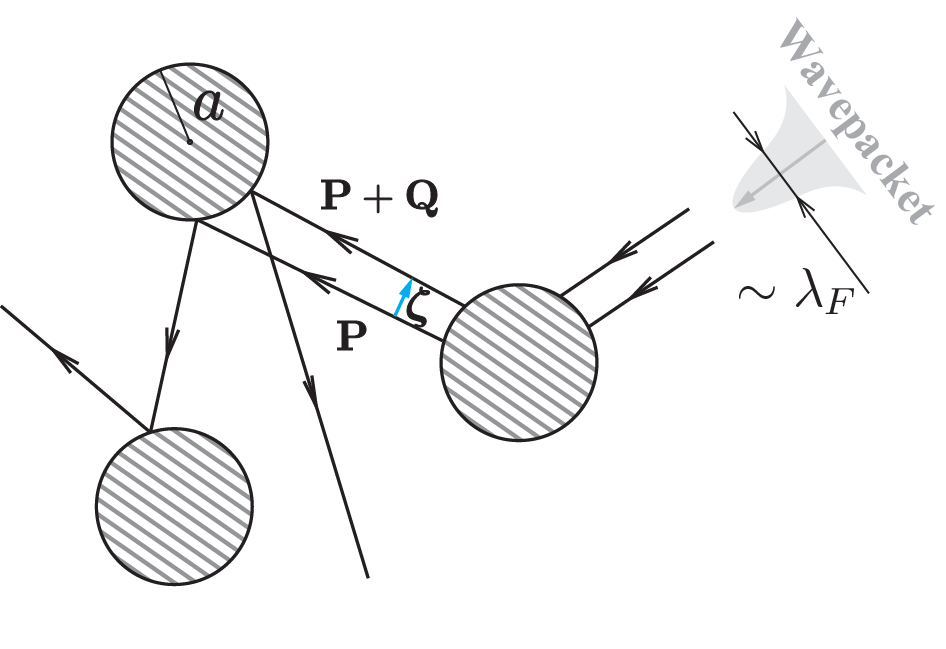}
	\caption{\label{ImpurityChain}
	(Colour online) Classical trajectories of electrons scattered by impurities.}
\end{figure}

{\it Impurity scattering.} In a non-interacting system,
electron wavepackets move along classical trajectories at sufficiently short times.
When two classical electrons with parallel momenta and slightly different impact parameters collide with an impurity,
their momenta get scattered at slightly different angles, as shown in Fig.~\ref{ImpurityChain}.
During subsequent collisions and propagation between
impurities, the distance and the angle between the trajectories grow further. This leads to the exponential
divergence between trajectories, as was first found in Ref.~\onlinecite{LarkinOvchnnikov} and
as also confirmed by the calculations in this paper, so long as the distance between trajectories
remains smaller than the characteristic impurity size $a$.
As soon as the distance between the trajectories exceeds the impurity size, the classical motion of electrons
becomes uncorrelated and the OTOC ceases to grow exponentially.

The typical initial distance between the trajectories which contribute to the OTOC is
determined by functions $K^{\alpha\beta}$, Eq.~(\ref{KAlphaBeta}), at $t=0$ and, thus,
by the function $g$, Eq.~\eqref{gDef}, whose
width may be estimated as $\lambda_F=2\pi/p_F$ for momenta $\bp_1$ and $\bp_2$ near the Fermi surface.
As a result of impurity collisions, the functions $K^{\alpha\beta}$ gets broadened to the characteristic
width $a$ in time on the order of the Ehrenfest time $t_E=\lambda^{-1}\ln(p_F a)$.

When considering the OTOC evolution on distances $|\bR_1-\bR_2|\gg \lambda_F$,
it is possible to make the approximation
$g(\bR_1-\bR_2,\bp_1,\bp_2)\approx -(2\pi)^d\delta(\bp_1-\bp_2)\delta(\bR_1-\bR_2)f_{\bp_1}(1-f_{\bp_1})$.
It follows then from Eq.~(\ref{gZZ}) that
\begin{align}
	\left<[\hat P_z(t),\hat P_z(0)]^2\right>=\left<\left(\frac{\partial p_z(t)}{\partial z(0)}\right)^2
	\right>_{g, \text{dis}},
	\label{PviaZ}
\end{align}
where $p_z(t)$ is the momentum of an electron along a classical trajectory with given initial
conditions $(\bp,\br)$; $z(0)$ is the coordinate along the $z$-axis at $t=0$; and
the averaging on the right-hand side is carried out with respect to the impurity locations (dis) and
the initial momentum $\bp$ and coordinate $\br$ of a classically moving electron:
\begin{align}
	\left<\ldots\right>_g=-\int_{\bp,\br}\ldots f_\bp(1-f_\bp).
	\label{gAve}
\end{align}
Eqs.~(\ref{PviaZ}) and (\ref{gAve}) illustrate that the momentum-OTOC characterises
the sensitivity of electron momenta at time $t$ to the change of the initial coordinates.

To obtain the momentum divergence of the classical electron trajectories we introduce the separation $\bzeta$
between respective pieces of the trajectories, as shown in Fig.~\ref{ImpurityChain}. 
Such a separation vector is well-defined, so long as the trajectories are close; the respective pieces
between two consecutive collisions may then be considered almost parallel. 
By analysing the scattering on a single impurity and free propagation between the impurities, we derive
(see Appendix~\ref{Sec:ClassDiverge} for details)
the equations for the evolution of the average separation $\bzeta$, the momentum difference $\bQ$
between the electrons (see Fig.~\ref{ImpurityChain})
and the correlator $\langle\bQ\cdot\bzeta\rangle$ between the momentum difference $\bQ$
and vector $\bzeta$:
\begin{align}
 \left\{
 \begin{array}{c}
  \frac{d}{dt}\left<Q^2\right>=\frac{4p_F^2\lambda^3}{v_F^2}\left<\zeta^2\right>,\\
  \frac{d}{dt}\left<\bzeta^2\right>=\frac{2v_F}{p_F}\left<\bzeta\cdot\bQ\right>,\\
  \frac{d}{dt}\left<\bzeta\cdot\bQ\right>=\frac{v_F}{p_F}\left<Q^2\right>,
 \end{array}
 \right.
 \label{SysQZeta}
\end{align}
where the averaging $\langle\ldots\rangle$ is carried out with respect to the impurity locations;
we assumed that the electron momenta are close to the Fermi momentum $p_F$ and have velocities $v_F$;
\begin{align}
	\lambda
	=\left(\frac{n_{\text{imp}} v_F^3}{4(d-1)}
	\int\left[\left(\frac{d\phi}{d\rho}\right)^2
	+(d-2)\frac{\sin^2\phi}{\rho^2}\right]S_{d-1}\rho^{d-2} d\rho\right)^\frac{1}{3}
	\label{Lambda}
\end{align}
is the classical Lyapunov exponent (for $d=3$,
the same leading rate of exponential divergence between trajectory pairs
was reported in Ref.~\onlinecite{LarkinOvchnnikov}); $n_{\text{imp}}$ is the impurity concentration;
$\phi(\rho)$ is the angle of scattering on a single impurity as a function of the impact parameter $\rho$,
and $S_{d-1}$ is the area of a unit sphere in a $d-1$-dimensional space (in this paper we consider $d>1$).
According to Eqs.~(\ref{SysQZeta}), the average momentum difference $\bQ(t)$
between two trajectories with the same initial momentum $p_F$ and separation $\zeta_0$
is given by
\begin{align}
	\left<Q^2(t)\right>=\frac{2}{3}\left(\frac{p_F \lambda\zeta_0}{v_F}\right)^2
	\left[	e^{2\lambda t}
	+2	e^{-{\lambda t}}\cos\left(\lambda t\sqrt{3}-\frac{2\pi}{3}\right)
	\right].
	\label{Q2growth}
\end{align}

{\it OTOC in a non-interacting system.}
For sufficiently long times, exceeding the transport scattering time
$\tau_{\text{tr}}=\left\{ v_Fn_{\text{imp}}
\int S_{d-1}\rho^{d-2}\left[1-\cos\phi(\rho)\right]d\rho\right\}^{-1}$,
the quasiparticle momentum $\bp(t)$ is uncorrelated with its initial direction, however,
there are still correlations between close pairs of trajectories which contribute to the
OTOC growth.
Using Eqs.~(\ref{PviaZ}), (\ref{gAve}) and (\ref{Q2growth}), we find 
the OTOC growth in a weakly disordered
non-interacting system:
\begin{align}
	F(t)
	=-\frac{2(d-1)}{3d^2}\frac{p_F^2 \lambda^2\nu_F}{v_F^2}V\cdot Te^{2\lambda t},
	\label{FwithT}
\end{align}
where we kept only the leading
exponentially growing contribution;
$\nu_F$ is the density of states at the Fermi surface in the $d$-dimensional conductor
under consideration; $V$ is the volume of the system; and we assume that the temperature $T$
is sufficiently high, $T\gg \lambda$. For very low temperatures, $T\ll \lambda$, the
OTOC is given by Eq.~\eqref{FwithT} with $T$ replaced by a constant of order $\lambda$.

{\it Electrons far from the Fermi surface.}
So far we considered quasiparticles near the Fermi surface.
We note, however, that quasiparticles far from the Fermi surface may yield significant contributions to the OTOC
in a non-interacting system.
Although their concentration is exponentially suppressed,
$\propto e^{-|E-E_F|/T}$, their contribution to the OTOC grows exponentially with velocity,
$\propto e^{\text{const}\cdot vt}$.
Strictly speaking, the OTOC in a non-interacting system is dominated by the quasiparticles with 
the largest velocity in the band.
However, in the presence of interaction, the inelastic relaxation rate grows rapidly away from
the Fermi surface, $1/\tau(E)\propto |E-E_F|^2$ for $|E-E_F|\gg T$, which strongly suppresses the lifetime
of excitations with high energies and their contributions to the OTOC.
In this paper we, therefore, neglect such high-energy excitations and focus on the quasiparticles near
the Fermi surface.

{\it Effect of interaction on OTOCs.} 
To the lowest order in the interaction strength (see Appendix~\ref{KineticEq}),
we derive 
the evolution of the correlation functions $K^{\alpha\beta}$,
Eq.~(\ref{KAlphaBeta}), between impurity collisions 
for small momentum difference $\bp_1-\bp_2$ in the form
\begin{align}
	\partial_t K^{\alpha\beta}(\bp_1,\bp_2)
	=-\left(\frac{1}{\tau_\alpha(\bp_1)}+\frac{1}{\tau_\beta(\bp_2)}\right)K^{\alpha\beta}(\bp_1,\bp_2)
	\nonumber \\
	+\frac{1}{\tau_{\bar\alpha}(\bp_1)}K^{\bar\alpha\beta}(\bp_1,\bp_2)
	+\frac{1}{\tau_{\bar\beta}(\bp_2)}K^{\alpha\bar\beta}(\bp_1,\bp_2),
	\label{InelEquation}
\end{align}
where we omitted the coordinate arguments of the functions $K^{\alpha\beta}$;
the indices $\alpha$ and $\beta$ again label electrons
and holes ($\alpha,\beta=e,h$) and we have introduced the electron scattering rate
\begin{align}
	\frac{1}{\tau_e(\bp)}=2\pi\int_{\bp^\prime,\bq}
	(1-f_{\bp^\prime})(1-f_{\bp+\bq})f_{\bp^\prime+\bq}|w(\bq)|^2
	\nonumber\\
	\delta(\xi_\bp+\xi_{\bp^\prime+\bq}-\xi_{\bp+\bq}-\xi_{\bp^\prime})
	\label{Erate}
\end{align}
and a similar hole scattering rate given by Eq.~(\ref{Erate}) with the replacements
$f\rightarrow 1-f$. For excitations on the Fermi surface and for short-range interactions
under consideration,
$1/\tau_e=1/\tau_h=1/\left[2\tau(T)\right]=T^2\cdot C k_F^{2d-3}\left[\int_\br |w(\br)|\right]^2/v_F^3 $,
where the constant $C\sim 1$ depends on the space dimensionality and the details of the Fermi surface.

Equation~(\ref{InelEquation}) may be understood qualitatively as follows. 
When two quasiparticles $\alpha$ and $\beta$ propagate along
close trajectories with close momenta $\bp_1$ and $\bp_2$,
each of them may get inelastically scattered, with
rates $1/\tau_\alpha$ and $1/\tau_\beta$, 
as reflected by the first term on the right-hand side of Eq.~(\ref{InelEquation}). 
In the spirit of Ref.~\onlinecite{Syzranov:OTOCdot}, quasiparticle states with significantly different
momenta may be considered as an external bath.
If one electron in a pair gets scattered, its momentum changes significantly
due to the short-range nature of the interaction, further motion of these electrons
is uncorrelated, and they no longer contribute to
the exponential growth of the OTOC.
There are also reverse processes, described by the last two terms in Eq.~(\ref{InelEquation});
a pair of electrons with close momenta may be created by inelastic processes; the respective 
contributions to $K^{ee}$, for example, require the existence of holes with momenta under consideration
and are thus proportional to $K^{eh,he}$.  We note also that we assume short-range interactions,
with radius $r_0\ll |\bR_1-\bR_2|$, which do not lead to direct interaction
between electrons on close trajectories. Depending on whether or not the interaction radius $r_0$
is smaller than $\lambda_F$, Eqs.~(\ref{InelEquation}) apply to the entire OTOC evolution or to
its later stages; in the latter case one has to use other effective initial conditions in place
of (\ref{gZZ}).

Eqs.~(\ref{InelEquation}) describe the evolution of the correlation functions $K^{\alpha\beta}$ 
between impurity collisions and indicate that distributions with
$K^{\alpha\alpha}\propto-K^{\alpha\bar{\alpha}}$ for all $\alpha$,
as corresponds to the initial conditions (\ref{gZZ}), relax with the rate $1/\tau_e+1/\tau_h=1/\tau(T)$,
where we have taken into account that electrons and holes have equal relaxation rates $1/\tau(T)$ on
the Fermi surface.
Thus, inelastic scattering suppresses correlations between pairs of electrons with close momenta
and leads to the 
exponential suppression of the OTOC as described by the 
second term in the exponent in Eq.~(\ref{Main}). At large temperatures or interaction strengths, 
this suppression prevents exponential growth, and a chaotic system becomes non-chaotic.

\begin{figure}[t]
	\centering
	\includegraphics[width=0.35\textwidth]{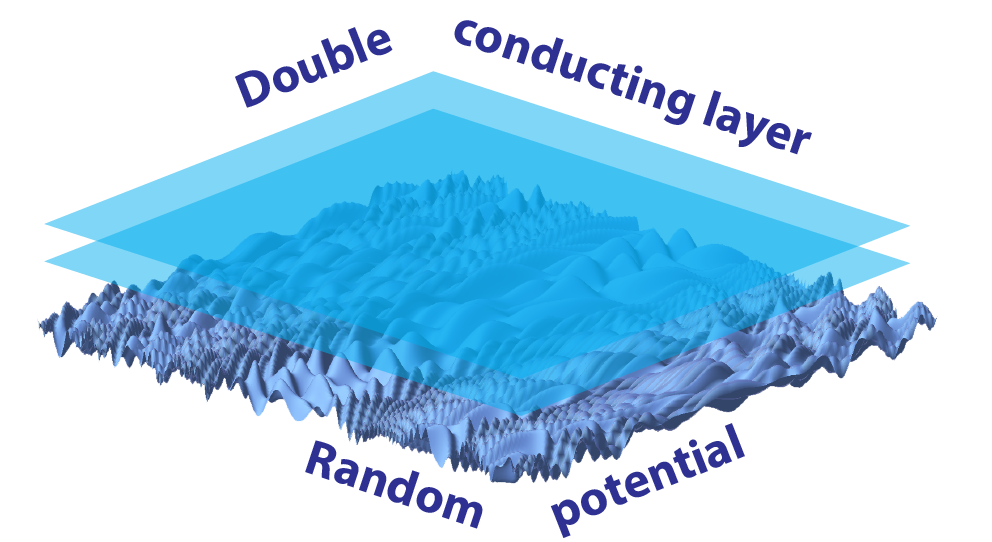}
	\caption{\label{Setup}
	(Colour online) Setup for measuring OTOCs; two layers of particles exposed to the same random potential.}
\end{figure}

{\it Relation between the symmetries of the system and the spectrum of Lyapunov exponents.}
The short-time behaviour of the correlators $K^{ee}$, $K^{eh}$ and $K^{hh}$, studied in this paper, describes
divergence between classical trajectories of, respectively, two electrons, an electron and a hole and two holes.
In the absence of an external magnetic field, close trajectories of two electrons diverge with exactly the
same rate (see Appendix~\ref{Sec:ClassDiverge} for a detailed derivation) as close trajectories of an electron and a hole.  
The rates are different, however, in the presence of a magnetic field; while the trajectories of pairs of electrons
remain close, the separation between electron and hole trajectories grows rapidly when a magnetic field is applied.
Thus, breaking time-reversal symmetry modifies qualitatively the structure of the set of Lyapunov exponents in the system.
We leave, however, for future studies a more rigorous investigation of the relation between the symmetries of the 
systems and the symmetries of the set of Lyapunov exponents.

{\it Relation to level statistics.} 
Systems
which exhibit chaotic dynamics in the classical limit, such as non-integrable billiards,
usually also
display Wigner-Dyson statistics of the energy levels~\cite{Efetov:book}. 
In particular, it was demonstrated recently\cite{Garcia:SYKtransition} that in a
generalised Sachdev-Ye-Kitaev model\cite{SachdevYe}, which is broadly used
as a toy model in the recent studies of quantum chaos, chaotic and non-chaotic behaviour correspond, respectively,
to the Wigner-Dyson and Poisson statistics of the many-body levels of the system.
We conjecture that the transition discussed in this paper
may be accompanied also by the change of quasiparticle level statistics characterised by the correlator
$R_2(\omega)=\left<\rho\left(E-\frac{\omega}{2}\right)\rho\left(E+\frac{\omega}{2}\right)\right>$,
where $\rho(E)=-\frac{1}{\pi V}\text{Im}\int_\br G^R(\br,\br,E)$ is the quasiparticle density of states,
and the averaging is carried out with respect to impurity locations.
We leave, however, a rigorous analysis of the relation between the exponential growth of OTOCs and
energy level statistics for future studies.

{\it Potential for experimental observation in double layers.} The evolution of the correlators
$K^{\alpha\beta}$ is similar to that of the correlations between two layers of particles exposed to the
same random potential, as shown in Fig.~\ref{Setup}, which suggests a way for experimental observations
of the OTOC and the transition between chaotic and non-chaotic behaviour
by observing momentum correlations between the two layers of, e.g., ultracold atoms or molecules.
Indeed, as it has been demonstrated in Ref.~\onlinecite{Syzranov:OTOCdot}, the OTOCs in a system
coupled to a classical environment may be mapped onto the evolution of two copies of that system
coupled to the same environment.
Initial correlations between
the particles in the layers may be induced, for example, by switching on attractive interactions between the layers
for a short time.

{\it Conclusion.} We have demonstrated that a weakly disordered metal with short-range interactions
displays a transition between quantum-chaotic and non-chaotic dynamics, identified through
the OTOC behaviour, when changing the temperature
or interaction strength. 
We conjecture that the transition is accompanied by a change in the level statistics
of the system.
Natural other future research directions include analysing other models
of interaction, quasiparticle dispersion, interplay with weak-localisation effects, etc.
Also, we expect our results to hold qualitatively if the
inelastic scattering comes from phonons or other types of external baths instead of electron-electron interactions,
because such a bath may suppress correlations between electron trajectories similarly to interactions.

Another question, which deserves a separate investigation, is the effect of rare events on the quantum chaotic
dynamics. Exponentially-rare fluctuations of the impurity density in a disordered material may lead to
sparse regions with large values of the Lyapunov exponent, which may affect the exponential
growth of OTOCs on sufficiently short times scales.
A detailed analysis of the effect of such rare events on the transition discussed in this paper
and on other aspects of quantum chaos will be presented elsewhere.

\begin{acknowledgments}
We acknowledge useful discussions with J.~Maldacena and B.~Swingle.
Our research was supported financially by NSF-DMR 1613029,
US-ARO contract No. W911NF1310172 (SVS), the U.S. Department of Energy contract
BES-DESC0001911, Simons Foundation (VMG) and  the 
Committee on Research at the University of California, Santa Cruz. AVG and SVS were also supported by 
NSF QIS, AFOSR, NSF PFC at JQI, ARO MURI, ARO and ARL CDQI.
\end{acknowledgments}


%

\newpage


\newpage

\onecolumngrid
\newpage

\appendix



\section{Classical divergence of trajectories}

\renewcommand{\theequation}{A\arabic{equation}}

\label{Sec:ClassDiverge}

\begin{figure}[h!]
	\centering
	\includegraphics[width=0.7\linewidth]{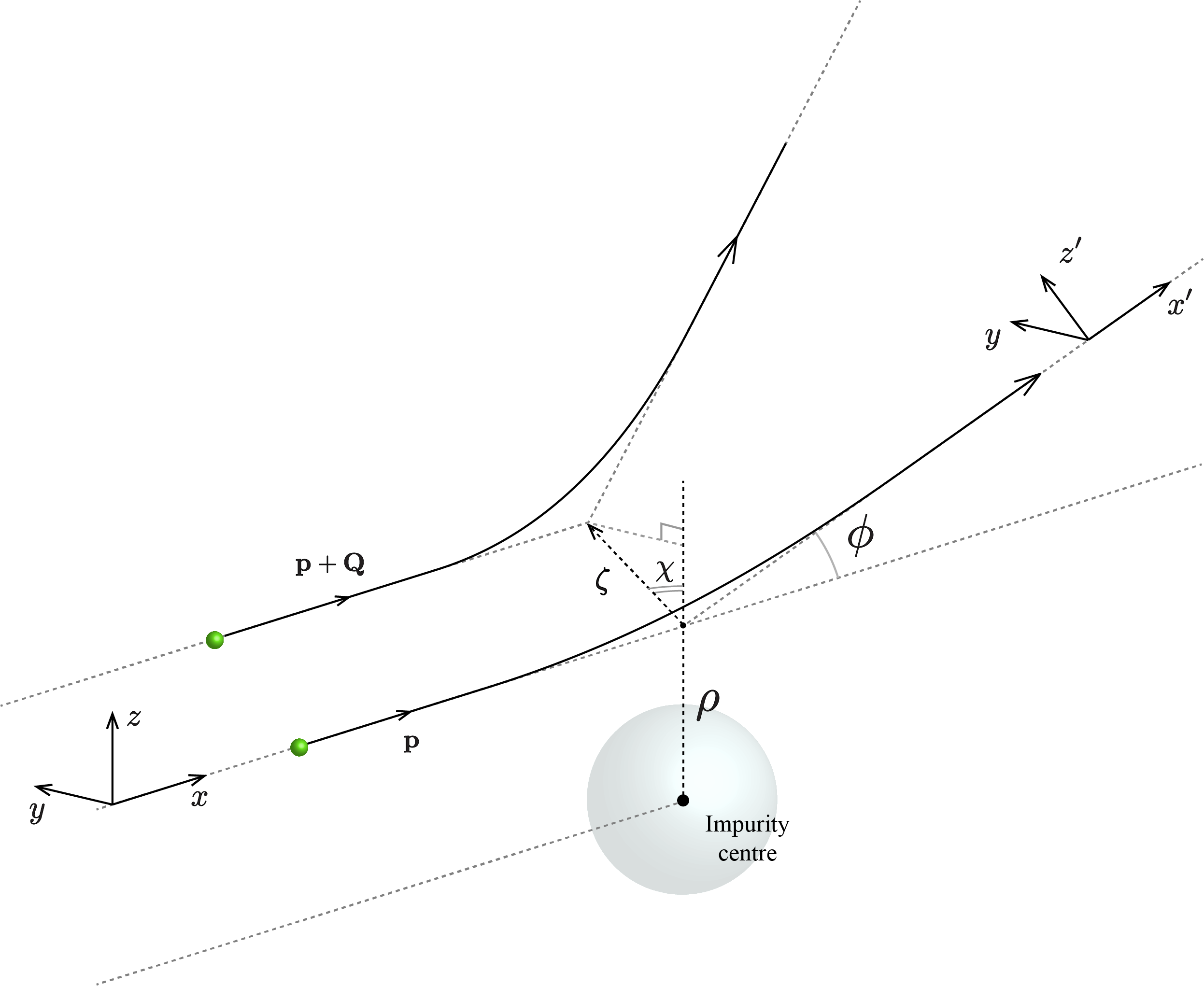}
	\caption{
	\label{fig:impurityscattering} Two trajectories of a particle scattering on an impurity. The initial directions
	velocities of the particle on the trajectories are almost parallel, corresponding to the momenta $\bp$ and $\bp+\bQ$;
	the initial distance between the trajectories is given by the vector $\zeta$ (while the differences between
	the trajectories are assumed to be very small, they are exaggerated in the figure). 
	}
\end{figure}

In this section we consider the divergence between the trajectories of a classical particle scattered off impurities.
We consider two trajectories which are almost parallel and characterise the distance between them by a small vector $\bzeta$,
almost orthogonal to the trajectories, as shown in Fig.~\ref{fig:impurityscattering}. The momenta of the particle
on the trajectories have the same absolute value but different directions; the momentum difference is characterised by a small
vector $\bQ$.
In what immediately follows we consider the transformation of the parameters $\bzeta$ and $\bQ$ when scattering on an
impurity. 
We assume that impurity collisions are rare and that the impurity potential is sufficiently short-ranged, which allows
us to 
consider scattering off each impurity in the system independently and
consider the trajectories as straight lines between scattering events.

We introduce the coordinate reference frame for the particle before the collision with an impurity, where the $x$ axis directed
along one of the trajectories far before the collision and the $z$ axis directed along the 
direction of the shortest distance from the centre of the impurity potential to the direction of the initial motion
of the particle, as shown in Fig.~\ref{fig:impurityscattering}.
It is convenient to introduce also the reference frame $x^\prime y^\prime z^\prime$ for characterizing the particles'
parameters after the collision, with the axis $x^\prime$ directed along the velocity (on the same trajectory)
after the collision and the axis $z^\prime$ lying in the $xz$ plane (the entire trajectory lies in the same plane
during the entire collision).
Due to the distance $\zeta$ between the trajectories, they hit the impurity with slightly different impact parameters:
$\rho$ and $\rho+\zeta\cos\chi$, where $\chi$ is the angle between the plane of one of the trajectories and vector $\bzeta$.
This gives a contribution to the difference between the directions after the collision, equivalent to a rotation around the $y^\prime$
axis by the angle $\frac{d\phi}{d\rho}\zeta\cos\chi$, where $\phi=\phi(\rho)$ is the scattering angle (as a function of the impact
parameter $\rho$). Taking into account that the particle momentum is $p$, such a rotation leads to the extra
contribution $\frac{d\phi}{d\rho}\cdot \zeta \cos\chi \cdot p$ to the momentum difference after the collision along the $z^\prime$ axis. 
The planes of the trajectories are rotated along the $x$ axis by a small angle of $\frac{\zeta\sin\chi}{\rho}$, which leads
to the contribution $\frac{1}{\rho}\cdot\zeta \sin\chi \cdot p\sin\phi$ to the 
momentum difference along the $y^\prime$ ($y$) direction after the collision.

Summing up all the contributions to the momentum difference, the transformation of the vector $\bQ$, taking into
account the rotation of the coordinate frame, is given by
\begin{align}
	\left\{
	\begin{array}{c}
	 Q_{z^\prime}=Q_z+\frac{d\phi}{d\rho}\cdot \zeta \cos\chi \cdot p+ \cO(\zeta^2,Q^2),\\
	 Q_{y^\prime}=Q_y+\frac{1}{\rho}\cdot\zeta \sin\chi \cdot p\sin\phi + \cO(\zeta^2,Q^2).
	\end{array}
	\right.
	\label{Qtransform}
\end{align}
In Eq.~(\ref{Qtransform}) we have taken into account that for $\zeta=0$ 
adding a small transverse momentum difference $\bQ$
is equivalent to rotating the trajectory around the centre of the impurity potential and, therefore,
$Q_{z^\prime}=Q_z$ and $Q_{y^\prime}=Q_y$ for $\zeta=0$.

Because the impurities are located randomly, the angle $\chi$ is a random variable distributed uniformly 
between $0$ and $2\pi$. Performing the averaging with respect to the angle $\chi$ and using Eqs.~(\ref{Qtransform}),
we arrive at
\begin{align}
	\left<(Q_{y^\prime}-Q_y)^2\right>_\chi+\left<(Q_{z^\prime}-Q_z)^2\right>_\chi
	=\frac{1}{2}\left(\frac{d\phi}{d\rho}\right)^2\zeta^2p^2+\frac{1}{2}\left(\frac{1}{\rho}\right)^2\zeta^2 p^2\sin^2\phi.
	\label{QtransformAngleAverage}
\end{align}
Performing the averaging of Eq.~(\ref{QtransformAngleAverage}) with respect to the impact parameter $\rho$ and with
respect to the vector $\bzeta$ (the distance between the trajectories) and introducing the
concentration $n_{\text{imp}}$ of the randomly located impurities, we obtain the growth rate of the amplitude of the
momentum difference $\bQ$:
\begin{align}
	\frac{d\left<\bQ^2\right>}{dt}=n_{\text{imp}}v\int \pi\rho
	\left[\left(\frac{d\phi}{d\rho}\right)^2+\frac{\sin^2\phi}{\rho^2}\right]d\rho\cdot\langle\bzeta^2\rangle,
	\label{Q2growth}
\end{align} 
where $v$ is the velocity of the particle long before and after the collision (identical for both trajectories).    

Taking into account the rotation of the coordinate frame, 
the separation $\bzeta$ between the trajectories does not change during the collision to the leading order in the small parameters
$\bzeta$ and $\bQ$. However, because the particle has different momenta on the two trajectories, the separation changes between 
the events of scattering on different impurities, with the time dependency described by the equation
\begin{align}
	\frac{d\left<\bzeta^2\right>}{dt}\equiv 2\left<\dot{\bzeta}\cdot\bzeta\right>= 
	\frac{2v}{p}\left<\bzeta\cdot\bQ\right>,
	\label{Zeta2growth}
\end{align}
where $\langle\ldots\rangle$ is the averaging with respect to the impurity locations.

In order to describe consistently the growth of the average distance and momentum separation between the trajectories,
Eqs.~(\ref{Q2growth}) and (\ref{Zeta2growth}) have to be complemented by the equation for the rate of change of 
the quantity
$\left<\bzeta\cdot\bQ\right>$.
Between collisions with impurities  $\dot \bQ=0$, and 
the separation $\bzeta$ between trajectories grows as $\dot \bzeta= \frac{v}{p}\bQ $.
Because the average elastic scattering time significantly exceeds the characteristic time
of colliding with one impurity, one may neglect the change of $\zeta$ during each collision compared
to that between collisions.
The change of the momentum difference $\bQ$ during a collision is given by Eq.~(\ref{Qtransform}).
Adding up the contributions to the rate of change of $\left<\bzeta\cdot\bQ\right>$ from both between and
during impurity collisions and performing the averaging with respect to the impurity locations,
we arrive at
\begin{subequations}
\begin{align}
	\frac{d}{dt}\langle\bzeta\cdot\bQ\rangle
	=\frac{v}{p}\left<\bQ^2\right>
	+pn_{\text{imp}}v \int \pi\rho
	\left(\frac{d\phi}{d\rho}+\frac{1}{\rho}\sin\phi\right)d\rho \cdot \langle\bzeta^2\rangle.
	\label{bZetaQChange}
\end{align}
Equations (\ref{Qtransform}), (\ref{Zeta2growth}) and (\ref{bZetaQChange}) constitute a complete system of equations
which govern the evolution of the variance of spatial and momentum separation between trajectories in a system of particles
scattered off randomly located impurities.

In each collision with an impurity, the momentum difference $Q$ changes by the typical value
$\delta Q\sim  p\zeta/a$, where $a$ is the characteristic radius of the impurity potential. As a result,
after collisions with impurities, the momentum difference becomes sufficiently large, 
$Q\gtrsim p\zeta/a$, and the first term in the right-hand side of Eq.~(\ref{bZetaQChange}), which is on the order 
$\frac{pv\zeta^2}{a^2}$, sufficiently exceeds the second term, which may be
estimated as $\sim pn_{\text{imp}}va\zeta^2$, taking into account the sparseness of the impurities ($na^3\ll1$).
Under this condition, Eq.~(\ref{bZetaQChange}) may be simplified as
\begin{align}
\frac{d}{dt}\langle\bzeta\cdot\bQ\rangle
=\frac{v}{p}\left<\bQ^2\right>.
\label{bZetaQChangeSimplified}
\end{align}
\end{subequations}
Equations \eqref{Q2growth}, \eqref{Zeta2growth} and \eqref{bZetaQChange} match the systems of equations \eqref{SysQZeta}
with the Lyapunov exponent given by Eq.~\eqref{Lambda} for $d=3$. The derivations of

\subsection*{Single-particle OTOC growth}

In the previous subsection we demonstrated that the dynamics of the 
momentum difference $\bQ$ and of the
separation $\bzeta$ between close classical trajectories are described by the system of equations
\begin{align}
\left\{
\begin{array}{c}
\frac{d}{dt}\left<Q^2\right>=\frac{4p^2\lambda^3}{v^2}\left<\zeta^2\right>,\\
\frac{d}{dt}\left<\bzeta^2\right>=\frac{2v}{p}\left<\bzeta\cdot\bQ\right>,\\
\frac{d}{dt}\left<\bzeta\cdot\bQ\right>=\frac{v}{p}\left<Q^2\right>.
\end{array}
\right.
\end{align}
The general solution for the average divergence $\left<\bQ^2(t)\right>$ of momenta is given by 
\begin{align}
	\left<\bQ^2(t)\right>=
	A_1 \exp\left(2\lambda t\right)
	+A_2 \exp\left(-\lambda t+i\lambda t\sqrt{3}\right)
	+A_3 \exp\left(-\lambda t-i\lambda t\sqrt{3}\right),
	\label{GenSolQ}
\end{align}
where $A_1$, $A_2$ and $A_3$ are constants determined by the initial conditions.
Two trajectories with the initial separation $\zeta_0$ and the same momenta correspond to the initial conditions
\begin{align}
	\left<\bQ^2(0)\right>=0,\quad \left<\bQ(0)\cdot\bzeta(0)\right>=0,
	\quad \left<\bzeta^2(0)\right>=\zeta_0^2,
\end{align}
which, together with Eq.~\ref{GenSolQ}, gives
\begin{align}
	\left<\bQ^2(t)\right>=
	\frac{2}{3}
	\left(\frac{p_F \zeta_0\lambda}{v_F}\right)^2
	\left[e^{2\lambda t}+e^{-\lambda t}\cos\left(\lambda t\sqrt{3}-\frac{2\pi}{3}\right)\right].
\end{align}


\subsection*{Reducing OTOC to the divergence of single-particle trajectories}

Using the function \eqref{KAlphaBeta}, introduced in the main text, the OTOC of the total momentum of
the electrons in a disordered metal may be represented in the form  
\begin{align}
F(t)=
&\int_{\bR_1,\bR_2,\bp_1,\bp_2} p_{1z}p_{2z}\: K^{ee}(\bp_1,\bR_{1};\bp_2,\bR_{2};t).
\label{FviaK}
\end{align}
The function $K^{ee}(\bp_1,\bR_{1};\bp_2,\bR_{2};t)$, introduced here, is an out-of-time-order correlator whose evolution
is similar to the product $f(\bp_1,\bR_{1})f(\bp_2,\bR_{2})$ of electron distribution functions in phase space.
In particular, for non-interacting particles, the evolution of function
$K^{ee}(\bp_1,\bR_{1};\bp_2,\bR_{2};t)$ and the product of the distribution functions are governed by the same kinetic 
equations.   

At small time $t$, when the propagation of electron wavepackets may be considered as that of classical particles,
the correlator $K^{ee}(\bp_1,\bR_{1};\bp_2,\bR_{2};t)$ is given by
\begin{align}
	K^{ee}(\bp_1,\bR_{1};\bp_2,\bR_{2};t)=
	\int_{\bR_1(0),\bR_2(0),\bp_1(0),\bp_2(0)}
	\delta\left\{\bp_1-\tilde{\bp}_1\left[\bp_1(0),\bR_{1}(0),t\right]\right\}
	\delta\left\{\bp_2-\tilde{\bp}_2\left[\bp_1(0),\bR_{1}(0),t\right]\right\}
	\nonumber\\
	\delta\left\{\bR_1-\tilde{\bR}_1\left[\bp_2(0),\bR_{2}(0),t\right]\right\}
	\delta\left\{\bR_2-\tilde{\bR}_2\left[\bp_2(0),\bR_{2}(0),t\right]\right\}
	\nonumber\\
		K^{ee}\left[\bp_1(0),\bR_{1}(0);\bp_2(0),\bR_{2}(0);0\right],
		\label{KGenEvolution}
\end{align}
where $\tilde{\bp}\left[\bp(0),\bR(0),t\right]$ and $\tilde{\bR}\left[\bp(0),\bR(0),t\right]$ are the momentum and coordinate along a classical trajectory
with initial conditions $[\bp(0),\bR(0)]$. From Eq.~\eqref{KGenEvolution} it follows that the OTOC~\eqref{FviaK}
is given by
\begin{align}
	F(t)=
	\int_{\bR_1(0),\bR_2(0),\bp_1(0),\bp_2(0)}
	\tilde{\bp}_1\left[\bp_1(0),\bR_{1}(0),t\right]
	\tilde{\bp}_2\left[\bp_1(0),\bR_{1}(0),t\right]
	K^{ee}\left[\bp_1(0),\bR_{1}(0);\bp_2(0),\bR_{2}(0);0\right].
	\label{FKgenericRelation}
\end{align}

The correlator $K^{ee}\left[\bp_1(0),\bR_{1}(0);\bp_2(0),\bR_{2}(0);0\right]$ at the initial time $t=0$ is a sharply
peaked function of the differences $|\bR_{1}(0)-\bR_{2}(0)|$ and $|\bp_1(0)-\bp_2(0)|$ of its arguments. As follows from 
the definition \eqref{KAlphaBeta}
of the function $K^{ee}(\bp_1,\bR_{1};\bp_2,\bR_{2};t)$, at $t=0$ it may be represented in the form
\begin{align}
	K^{ee}(\bp_1,\bR_{1};\bp_2,\bR_{2};0)=\partial_{Z_1}\partial_{Z_2}g(\bR_1-\bR_2,\bp_1-\bp_2),
	\label{KgRelation}
\end{align}
where the function $g(\bR_1-\bR_2,\bp_1-\bp_2)$ is given by Eq.~\eqref{gDef} and may be approximated by a delta-function in
phase space
\begin{align}
	g(\bR_1-\bR_2,\bp_1-\bp_2)\approx -(2\pi)^d\delta(\bp_1-\bp_2)\delta(\bR_1-\bR_2)f_{\bp_1}(1-f_{\bp_1})
	\label{gInit}
\end{align}
when dynamics on large scales $|\bR_{1}-\bR_{2}|\gg\lambda_F$ is considered, where $\lambda_F$ is the Fermi length in the 
systems under consideration. 

Using equations \eqref{gInit}, \eqref{KgRelation} and \eqref{FKgenericRelation} gives
\begin{align}
F(t)=\left<\left(\frac{\partial p_z(t)}{\partial z(0)}\right)^2
\right>_{g, \text{dis}},
\label{FrepeatClassRelation}
\end{align}
with $\left<\ldots\right>_g=-\int_{\bp,\br}\ldots f_\bp(1-f_\bp)$. Equation \eqref{FrepeatClassRelation}
relates the OTOC~\eqref{Main} of quantum operators in a systems of quantum non-interacting particles with the sensitivity
of classical trajectories of those particles to their initial coordinates $z(0)$.


\section{Correlators in the presence of interactions}

\renewcommand{\theequation}{B\arabic{equation}}

\label{KineticEq}

In this section, we outline a phenomenological derivation of the kinetic equation \eqref{InelEquation}; a detailed rigorous derivation
will be presented in Ref.~\onlinecite{Syzranov:OTOCdetail}.
 We focus on the correlator
$K^{ee}(\bp_1,\bR_1;\bp_2,\bR_2;t)$; the kinetic equations for the other correlators may be derived similarly. Here, we assume that the momenta $\bp_1$
and $\bp_2$ are close to each other and treat the rest of the electrons, i.e. electrons with significantly different momenta, as 
an independent bath, in the spirit of Ref.~\onlinecite{Syzranov:OTOCdot}. The locations $\bR_{1}$ and $\bR_{2}$ are supposed to be sufficiently close to each other,
closer than the impurity size $a$, but further away from than the interaction radius and the Fermi wavelength. 
The function $K^{ee}(\bp_1,\bR_1;\bp_2,\bR_2;t)$ is assumed to be a smooth function of its coordinates.

Under these assumptions it is possible to introduce the basis of plane-wave-like states $\bp$ and $\bk$ in small vicinities of the locations $\bR_{1}$
and $\bR_{2}$ respectively and consider the evolution of the operator
\begin{align}
	\tilde{K}(\bp,\bk)\propto\left<\left[\ha_\bp^\dagger\ha_\bp,
	\hat P_z(0)\right]
	\left[\ha_\bk^\dagger\ha_\bk,
	\hat P_z(0)\right]\right>,
	\label{Kpseudo}
\end{align}
where $\ha_\bp^\dagger$ and $\ha_\bp$ are the creation an annihilation operators of the respective plane-wave states near locations $\bp$ and $\bk$.
Under the made assumptions, the correlator \eqref{Kpseudo} obeys the same kinetic equation as the correlator $K^{ee}(\bp_1,\bR_1;\bp_2,\bR_2;t)$. To the leading order
in the Hamiltonian of interactions
\begin{align}
	\cH_{\text{int}}=\frac{1}{V}\sum_{\bp,\bp^\prime,\bq}\ha^\dagger_{\bp^\prime+\bq}\ha^\dagger_\bp
	\ha_{\bp+\bq}\ha_{\bp^\prime}U(\bq),
	\label{Hint}
\end{align}
the evolution of the correlator is governed by the kinetic equation
\begin{align}
	\partial_t \tilde{K}(\bp,\bk)=
	&-\left<\left[\int_{-\infty}^t\left[\cH_{\text{int}}(t^\prime),\left[\cH_{\text{int}}(t),\ha_\bp^\dagger\ha_\bp\right]\right]dt^\prime\,,\hat P_z(0)\right]
	\left[\ha_\bk^\dagger\ha_\bk,
	\hat P_z(0)\right]\right>
	\nonumber\\
	&-\left<\left[\ha_\bp^\dagger\ha_\bp,\hat P_z(0)\right]
	\left[\int_{-\infty}^t\left[\cH_{\text{int}}(t^\prime),
	\left[\cH_{\text{int}}(t),\ha_\bk^\dagger\ha_\bk(t)\right]\right]dt^\prime\,,\hat P_z(0)\right]\right>
	\nonumber\\
	&-\left<\left[\left[\cH_{\text{int}}(t),\ha_\bp^\dagger\ha_\bp(t)\right],\hat P_z(0)\right]\left[
	\int_{-\infty}^t\left[\cH_{\text{int}}(t^\prime),\ha_\bk^\dagger\ha_\bk(t)\right]dt^\prime\,,\hat P_z(0)\right]\right>
	\nonumber\\
	&-\left<\left[\int_{-\infty}^t\left[\cH_{\text{int}}(t^\prime),\ha_\bp^\dagger\ha_\bp(t)\right]dt^\prime\,,\hat P_z(0)\right]
	\left[\left[\cH_{\text{int}}(t),\ha_\bk^\dagger\ha_\bk(t)\right],\hat P_z(0)\right]\right>.
	\label{PreKineticPert}
\end{align}
Equation \eqref{PreKineticPert} is similar to the kinetic for an OTOC derived in Ref.~\onlinecite{Syzranov:OTOCdot} for a system with discrete energy levels.
The first two terms in the right-hand side of Eq.~\ref{PreKineticPert} describe the change of the correlator
due to the interactions of each quasiparticle with the Fermi sea, while the last two terms describe
correlations between different phase trajectories of the quasiparticles due to the interactions.
Because we consider interactions with a short radius, exceeded by the distance $|\bR_{1}-\bR_{2}|$,
the effect of the last two terms may be neglected.

Under the made assumptions, the correlators in Eq.~\eqref{PreKineticPert} may be represented as products of independent
correlators or operators of momentum states near $\bp$ and $\bk$ and correlators of momenta 
far from $\bp$ and $\bk$. 
From Eq.~\eqref{PreKineticPert} we obtain straightforwardly, using the form~\eqref{Hint} of the interaction Hamiltonian,
\begin{align}
	\partial_t K^{ee}\left(\bp_1,\bp_2\right)
	=-\left[\frac{1}{\tau_e(\bp_1)}+\frac{1}{\tau_e(\bp_2)}\right]K^{ee}\left(\bp_1,\bp_2\right)
	+\frac{1}{\tau_h(\bp_2)}K^{eh}\left(\bp_1,\bp_2\right)
	+\frac{1}{\tau_h(\bp_1)}K^{he}\left(\bp_1,\bp_2\right),
\end{align}
where $\tau_e(\bp_1)$ is the electron scattering rate given by Eq.~\eqref{Erate} and $\tau_h(\bp_1)$
is the hole scattering rate given by the same equation with the replacement $f\rightarrow 1-f$.

\subsubsection*{Eigenmodes of the correlators}

In what follows, we consider for simplicity short-range interactions
and quasiparticle momenta close to the Fermi surface.
The scattering times $\tau_e(\bp)=\tau_h(\bp)\equiv 1/(2\tau)\propto T^2$ are then
determined by collisions with other quasiparticles and are identical for electrons and holes and
independent of the direction of momentum.
The kinetic equations \eqref{InelEquation} are a system of four linear eqautions for the correlators
$K^{\alpha\beta}(\bp_1,\bR_1;\bp_2,\bR_2;t)$. This system has four eigenvector given by
\begin{align}
	\left(
	\begin{array}{c}
	K^{ee}\\
	K^{eh}\\
	K^{he}\\
	K^{hh}
	\end{array}
	\right)\propto
	\left(
	\begin{array}{c}
	1\\
	1\\
	1\\
	1
	\end{array}
	\right),
		\left(
	\begin{array}{c}
	1\\
	0\\
	0\\
	-1
	\end{array}
	\right),
		\left(
	\begin{array}{c}
	0\\
	-1\\
	1\\
	0
	\end{array}
	\right),
		\left(
	\begin{array}{c}
	1\\
	-1\\
	-1\\
	1
	\end{array}
	\right)
	\label{Eigenvectors}
\end{align}
and corresponding respectively, to the relaxation rates of the correlators given by 
$0$, $\tau^{-1}$, $\tau^{-1}$ and $2\tau^{-1}$. 

The initial conditions for the correlators $K^{\alpha\beta}(\bp_1,\bR_1;\bp_2,\bR_2;t)$, given by Eq.~\eqref{KAlphaBeta},
match the fourth eigenvector \eqref{Eigenvectors}, corresponding to a decay rate of $2\tau^{-1}$.
When both impurity scattering and interactions are present, then under the made assumptions of the sort radius and weakness
of the interactions, the OTOC is given by
\begin{align}
	F(t)=F_0(t) \exp\left(-2t/\tau\right),
\end{align} 
where $F_0$ is the OTOC in a non-interacting system.


\begin{thebibliography}{31}%
	\makeatletter
	\providecommand \@ifxundefined [1]{%
		\@ifx{#1\undefined}
	}%
	\providecommand \@ifnum [1]{%
		\ifnum #1\expandafter \@firstoftwo
		\else \expandafter \@secondoftwo
		\fi
	}%
	\providecommand \@ifx [1]{%
		\ifx #1\expandafter \@firstoftwo
		\else \expandafter \@secondoftwo
		\fi
	}%
	\providecommand \natexlab [1]{#1}%
	\providecommand \enquote  [1]{``#1''}%
	\providecommand \bibnamefont  [1]{#1}%
	\providecommand \bibfnamefont [1]{#1}%
	\providecommand \citenamefont [1]{#1}%
	\providecommand \href@noop [0]{\@secondoftwo}%
	\providecommand \href [0]{\begingroup \@sanitize@url \@href}%
	\providecommand \@href[1]{\@@startlink{#1}\@@href}%
	\providecommand \@@href[1]{\endgroup#1\@@endlink}%
	\providecommand \@sanitize@url [0]{\catcode `\\12\catcode `\$12\catcode
		`\&12\catcode `\#12\catcode `\^12\catcode `\_12\catcode `\%12\relax}%
	\providecommand \@@startlink[1]{}%
	\providecommand \@@endlink[0]{}%
	\providecommand \url  [0]{\begingroup\@sanitize@url \@url }%
	\providecommand \@url [1]{\endgroup\@href {#1}{\urlprefix }}%
	\providecommand \urlprefix  [0]{URL }%
	\providecommand \Eprint [0]{\href }%
	\providecommand \doibase [0]{http://dx.doi.org/}%
	\providecommand \selectlanguage [0]{\@gobble}%
	\providecommand \bibinfo  [0]{\@secondoftwo}%
	\providecommand \bibfield  [0]{\@secondoftwo}%
	\providecommand \translation [1]{[#1]}%
	\providecommand \BibitemOpen [0]{}%
	\providecommand \bibitemStop [0]{}%
	\providecommand \bibitemNoStop [0]{.\EOS\space}%
	\providecommand \EOS [0]{\spacefactor3000\relax}%
	\providecommand \BibitemShut  [1]{\csname bibitem#1\endcsname}%
	\let\auto@bib@innerbib\@empty
	\bibitem [{\citenamefont {Efetov}(1999)}]{Efetov:book}%
	\BibitemOpen
	\bibfield  {author} {\bibinfo {author} {\bibfnamefont {K.~B.}\ \bibnamefont
			{Efetov}},\ }\href@noop {} {\emph {\bibinfo {title} {Supersymetry in Disorder
				and Chaos}}}\ (\bibinfo  {publisher} {Cambridge University Press},\ \bibinfo
	{address} {New York},\ \bibinfo {year} {1999})\BibitemShut {NoStop}%
	\bibitem [{\citenamefont {Larkin}\ and\ \citenamefont
		{Ovchinnikov}(1969)}]{LarkinOvchnnikov}%
	\BibitemOpen
	\bibfield  {author} {\bibinfo {author} {\bibfnamefont {A.}~\bibnamefont
			{Larkin}}\ and\ \bibinfo {author} {\bibfnamefont {Y.~N.}\ \bibnamefont
			{Ovchinnikov}},\ }\bibfield  {title} {\enquote {\bibinfo {title}
			{Quasiclassical method in the theory of superconductivity},}\ }\href@noop {}
	{\bibfield  {journal} {\bibinfo  {journal} {Sov. Phys. JETP}\ }\textbf
		{\bibinfo {volume} {28}},\ \bibinfo {pages} {960} (\bibinfo {year}
		{1969})}\BibitemShut {NoStop}%
	\bibitem [{\citenamefont {{Maldacena}}\ \emph {et~al.}(2016)\citenamefont
		{{Maldacena}}, \citenamefont {{Shenker}},\ and\ \citenamefont
		{{Stanford}}}]{Maldacena:bound}%
	\BibitemOpen
	\bibfield  {author} {\bibinfo {author} {\bibfnamefont {J.}~\bibnamefont
			{{Maldacena}}}, \bibinfo {author} {\bibfnamefont {S.~H.}\ \bibnamefont
			{{Shenker}}}, \ and\ \bibinfo {author} {\bibfnamefont {D.}~\bibnamefont
			{{Stanford}}},\ }\bibfield  {title} {\enquote {\bibinfo {title} {{A bound on
					chaos}},}\ }\href@noop {} {\bibfield  {journal} {\bibinfo  {journal} {JHEP}\
		}\textbf {\bibinfo {volume} {8}},\ \bibinfo {pages} {106} (\bibinfo {year}
		{2016})}\BibitemShut {NoStop}%
	\bibitem [{\citenamefont {Aleiner}\ \emph {et~al.}(2016)\citenamefont
		{Aleiner}, \citenamefont {Faoro},\ and\ \citenamefont
		{Ioffe}}]{AleinerFaoroIoffe}%
	\BibitemOpen
	\bibfield  {author} {\bibinfo {author} {\bibfnamefont {Igor~L.}\ \bibnamefont
			{Aleiner}}, \bibinfo {author} {\bibfnamefont {Lara}\ \bibnamefont {Faoro}}, \
		and\ \bibinfo {author} {\bibfnamefont {Lev~B.}\ \bibnamefont {Ioffe}},\
	}\href@noop {} {\enquote {\bibinfo {title} {Microscopic model of quantum
				butterfly effect: out-of-time-order correlators and traveling combustion
				waves},}\ } (\bibinfo {year} {2016}),\ \bibinfo {note}
	{arXiv:1609.01251}\BibitemShut {NoStop}%
	\bibitem [{\citenamefont {{Fan}}\ \emph {et~al.}(2017)\citenamefont {{Fan}},
		\citenamefont {{Zhang}}, \citenamefont {{Shen}},\ and\ \citenamefont
		{{Zhai}}}]{Fan:OTOCMBL}%
	\BibitemOpen
	\bibfield  {author} {\bibinfo {author} {\bibfnamefont {R.}~\bibnamefont
			{{Fan}}}, \bibinfo {author} {\bibfnamefont {P.}~\bibnamefont {{Zhang}}},
		\bibinfo {author} {\bibfnamefont {H.}~\bibnamefont {{Shen}}}, \ and\ \bibinfo
		{author} {\bibfnamefont {H.}~\bibnamefont {{Zhai}}},\ }\bibfield  {title}
	{\enquote {\bibinfo {title} {{Out-of-Time-Order Correlation for Many-Body
					Localization}},}\ }\href@noop {} {\bibfield  {journal} {\bibinfo  {journal}
			{Science Bull.}\ }\textbf {\bibinfo {volume} {62}},\ \bibinfo {pages} {707}
		(\bibinfo {year} {2017})}\BibitemShut {NoStop}%
	\bibitem [{\citenamefont {{Chen}}(2016)}]{Chen:MBLscrambling}%
	\BibitemOpen
	\bibfield  {author} {\bibinfo {author} {\bibfnamefont {Y.}~\bibnamefont
			{{Chen}}},\ }\bibfield  {title} {\enquote {\bibinfo {title} {Universal
				logarithmic scrambling in many body localization},}\ }\href@noop {}
	{\bibfield  {journal} {\bibinfo  {journal} {ArXiv e-prints}\ } (\bibinfo
		{year} {2016})},\ \Eprint {http://arxiv.org/abs/1608.02765} {arXiv:1608.02765
		[cond-mat.dis-nn]} \BibitemShut {NoStop}%
	\bibitem [{\citenamefont {{Swingle}}\ and\ \citenamefont
		{{Chowdhury}}(2017)}]{SwingleChowdhury:LocScrambling}%
	\BibitemOpen
	\bibfield  {author} {\bibinfo {author} {\bibfnamefont {B.}~\bibnamefont
			{{Swingle}}}\ and\ \bibinfo {author} {\bibfnamefont {D.}~\bibnamefont
			{{Chowdhury}}},\ }\bibfield  {title} {\enquote {\bibinfo {title} {{Slow
					scrambling in disordered quantum systems}},}\ }\href@noop {} {\bibfield
		{journal} {\bibinfo  {journal} {Phys. Rev. B}\ }\textbf {\bibinfo {volume}
			{95}},\ \bibinfo {pages} {060201} (\bibinfo {year} {2017})}\BibitemShut
	{NoStop}%
	\bibitem [{\citenamefont {{Huang}}\ \emph {et~al.}(2017)\citenamefont
		{{Huang}}, \citenamefont {{Zhang}},\ and\ \citenamefont
		{{Chen}}}]{HuangChen:OTOCMBL}%
	\BibitemOpen
	\bibfield  {author} {\bibinfo {author} {\bibfnamefont {Y.}~\bibnamefont
			{{Huang}}}, \bibinfo {author} {\bibfnamefont {Y.-L.}\ \bibnamefont
			{{Zhang}}}, \ and\ \bibinfo {author} {\bibfnamefont {X.}~\bibnamefont
			{{Chen}}},\ }\bibfield  {title} {\enquote {\bibinfo {title}
			{{Out-of-time-ordered correlators in many-body localized systems}},}\
	}\href@noop {} {\bibfield  {journal} {\bibinfo  {journal} {Ann. Phys.}\
		}\textbf {\bibinfo {volume} {529}},\ \bibinfo {pages} {1600318} (\bibinfo
		{year} {2017})}\BibitemShut {NoStop}%
	\bibitem [{\citenamefont {Rozenbaum}\ \emph {et~al.}(2017)\citenamefont
		{Rozenbaum}, \citenamefont {Ganeshan},\ and\ \citenamefont
		{Galitski}}]{Rosenbaum:rotor}%
	\BibitemOpen
	\bibfield  {author} {\bibinfo {author} {\bibfnamefont {Efim~B.}\ \bibnamefont
			{Rozenbaum}}, \bibinfo {author} {\bibfnamefont {Sriram}\ \bibnamefont
			{Ganeshan}}, \ and\ \bibinfo {author} {\bibfnamefont {Victor}\ \bibnamefont
			{Galitski}},\ }\bibfield  {title} {\enquote {\bibinfo {title} {Lyapunov
				exponent and out-of-time-ordered correlator's growth rate in a chaotic
				system},}\ }\href@noop {} {\bibfield  {journal} {\bibinfo  {journal} {Phys.
				Rev. Lett.}\ }\textbf {\bibinfo {volume} {118}},\ \bibinfo {pages} {086801}
		(\bibinfo {year} {2017})}\BibitemShut {NoStop}%
	\bibitem [{\citenamefont {Patel}\ \emph {et~al.}(2017)\citenamefont {Patel},
		\citenamefont {Chowdhury}, \citenamefont {Sachdev},\ and\ \citenamefont
		{Swingle}}]{PatelSachdev}%
	\BibitemOpen
	\bibfield  {author} {\bibinfo {author} {\bibfnamefont {Aavishkar~A.}\
			\bibnamefont {Patel}}, \bibinfo {author} {\bibfnamefont {Debanjan}\
			\bibnamefont {Chowdhury}}, \bibinfo {author} {\bibfnamefont {Subir}\
			\bibnamefont {Sachdev}}, \ and\ \bibinfo {author} {\bibfnamefont {Brian}\
			\bibnamefont {Swingle}},\ }\bibfield  {title} {\enquote {\bibinfo {title}
			{Quantum butterfly effect in weakly interacting diffusive metals},}\
	}\href@noop {} {\bibfield  {journal} {\bibinfo  {journal} {Phys. Rev. X}\
		}\textbf {\bibinfo {volume} {7}},\ \bibinfo {pages} {031047} (\bibinfo {year}
		{2017})}\BibitemShut {NoStop}%
	\bibitem [{\citenamefont {{Bagrets}}\ \emph {et~al.}(2017)\citenamefont
		{{Bagrets}}, \citenamefont {{Altland}},\ and\ \citenamefont
		{{Kamenev}}}]{BagretsKamenev}%
	\BibitemOpen
	\bibfield  {author} {\bibinfo {author} {\bibfnamefont {D.}~\bibnamefont
			{{Bagrets}}}, \bibinfo {author} {\bibfnamefont {A.}~\bibnamefont
			{{Altland}}}, \ and\ \bibinfo {author} {\bibfnamefont {A.}~\bibnamefont
			{{Kamenev}}},\ }\bibfield  {title} {\enquote {\bibinfo {title} {{Power-law
					out of time order correlation functions in the SYK model}},}\ }\href@noop {}
	{\bibfield  {journal} {\bibinfo  {journal} {Nuclear Physics B}\ }\textbf
		{\bibinfo {volume} {921}},\ \bibinfo {pages} {727--752} (\bibinfo {year}
		{2017})}\BibitemShut {NoStop}%
	\bibitem [{\citenamefont {{Werman}}\ \emph {et~al.}(2017)\citenamefont
		{{Werman}}, \citenamefont {{Kivelson}},\ and\ \citenamefont
		{{Berg}}}]{Kivelson:phonons}%
	\BibitemOpen
	\bibfield  {author} {\bibinfo {author} {\bibfnamefont {Y.}~\bibnamefont
			{{Werman}}}, \bibinfo {author} {\bibfnamefont {S.~A.}\ \bibnamefont
			{{Kivelson}}}, \ and\ \bibinfo {author} {\bibfnamefont {E.}~\bibnamefont
			{{Berg}}},\ }\bibfield  {title} {\enquote {\bibinfo {title} {{Quantum chaos
					in an electron-phonon bad metal}},}\ }\href@noop {} {\bibfield  {journal}
		{\bibinfo  {journal} {ArXiv e-prints}\ } (\bibinfo {year} {2017})},\ \Eprint
	{http://arxiv.org/abs/1705.07895} {arXiv:1705.07895 [cond-mat.str-el]}
	\BibitemShut {NoStop}%
	\bibitem [{\citenamefont {{Yao}}\ \emph {et~al.}(2016)\citenamefont {{Yao}},
		\citenamefont {{Grusdt}}, \citenamefont {{Swingle}}, \citenamefont {{Lukin}},
		\citenamefont {{Stamper-Kurn}}, \citenamefont {{Moore}},\ and\ \citenamefont
		{{Demler}}}]{Yao:2copies}%
	\BibitemOpen
	\bibfield  {author} {\bibinfo {author} {\bibfnamefont {N.~Y.}\ \bibnamefont
			{{Yao}}}, \bibinfo {author} {\bibfnamefont {F.}~\bibnamefont {{Grusdt}}},
		\bibinfo {author} {\bibfnamefont {B.}~\bibnamefont {{Swingle}}}, \bibinfo
		{author} {\bibfnamefont {M.~D.}\ \bibnamefont {{Lukin}}}, \bibinfo {author}
		{\bibfnamefont {D.~M.}\ \bibnamefont {{Stamper-Kurn}}}, \bibinfo {author}
		{\bibfnamefont {J.~E.}\ \bibnamefont {{Moore}}}, \ and\ \bibinfo {author}
		{\bibfnamefont {E.~A.}\ \bibnamefont {{Demler}}},\ }\bibfield  {title}
	{\enquote {\bibinfo {title} {{Interferometric Approach to Probing Fast
					Scrambling}},}\ }\href@noop {} {\bibfield  {journal} {\bibinfo  {journal}
			{ArXiv e-prints}\ } (\bibinfo {year} {2016})},\ \Eprint
	{http://arxiv.org/abs/1607.01801} {arXiv:1607.01801} \BibitemShut {NoStop}%
	\bibitem [{\citenamefont {{Bohrdt}}\ \emph {et~al.}(2017)\citenamefont
		{{Bohrdt}}, \citenamefont {{Mendl}}, \citenamefont {{Endres}},\ and\
		\citenamefont {{Knap}}}]{Knap:2copies}%
	\BibitemOpen
	\bibfield  {author} {\bibinfo {author} {\bibfnamefont {A.}~\bibnamefont
			{{Bohrdt}}}, \bibinfo {author} {\bibfnamefont {C.~B.}\ \bibnamefont
			{{Mendl}}}, \bibinfo {author} {\bibfnamefont {M.}~\bibnamefont {{Endres}}}, \
		and\ \bibinfo {author} {\bibfnamefont {M.}~\bibnamefont {{Knap}}},\
	}\bibfield  {title} {\enquote {\bibinfo {title} {{Scrambling and
					thermalization in a diffusive quantum many-body system}},}\ }\href@noop {}
	{\bibfield  {journal} {\bibinfo  {journal} {New J. Phys.}\ }\textbf {\bibinfo
			{volume} {19}},\ \bibinfo {pages} {063001} (\bibinfo {year}
		{2017})}\BibitemShut {NoStop}%
	\bibitem [{\citenamefont {{Swingle}}\ \emph {et~al.}(2016)\citenamefont
		{{Swingle}}, \citenamefont {{Bentsen}}, \citenamefont {{Schleier-Smith}},\
		and\ \citenamefont {{Hayden}}}]{Swingle:measurement}%
	\BibitemOpen
	\bibfield  {author} {\bibinfo {author} {\bibfnamefont {B.}~\bibnamefont
			{{Swingle}}}, \bibinfo {author} {\bibfnamefont {G.}~\bibnamefont
			{{Bentsen}}}, \bibinfo {author} {\bibfnamefont {M.}~\bibnamefont
			{{Schleier-Smith}}}, \ and\ \bibinfo {author} {\bibfnamefont
			{P.}~\bibnamefont {{Hayden}}},\ }\bibfield  {title} {\enquote {\bibinfo
			{title} {{Measuring the scrambling of quantum information}},}\ }\href@noop {}
	{\bibfield  {journal} {\bibinfo  {journal} {Phys. Rev. A}\ }\textbf {\bibinfo
			{volume} {94}},\ \bibinfo {pages} {040302} (\bibinfo {year}
		{2016})}\BibitemShut {NoStop}%
	\bibitem [{\citenamefont {Zhu}\ \emph {et~al.}(2016)\citenamefont {Zhu},
		\citenamefont {Hafezi},\ and\ \citenamefont
		{Grover}}]{ZhuHafeziGrover:measureClock}%
	\BibitemOpen
	\bibfield  {author} {\bibinfo {author} {\bibfnamefont {Guanyu}\ \bibnamefont
			{Zhu}}, \bibinfo {author} {\bibfnamefont {Mohammad}\ \bibnamefont {Hafezi}},
		\ and\ \bibinfo {author} {\bibfnamefont {Tarun}\ \bibnamefont {Grover}},\
	}\bibfield  {title} {\enquote {\bibinfo {title} {Measurement of many-body
				chaos using a quantum clock},}\ }\href@noop {} {\bibfield  {journal}
		{\bibinfo  {journal} {Phys. Rev. A}\ }\textbf {\bibinfo {volume} {94}},\
		\bibinfo {pages} {062329} (\bibinfo {year} {2016})}\BibitemShut {NoStop}%
	\bibitem [{\citenamefont {{Danshita}}\ \emph {et~al.}(2017)\citenamefont
		{{Danshita}}, \citenamefont {{Hanada}},\ and\ \citenamefont
		{{Tezuka}}}]{Danshita:SYKmeasurement}%
	\BibitemOpen
	\bibfield  {author} {\bibinfo {author} {\bibfnamefont {I.}~\bibnamefont
			{{Danshita}}}, \bibinfo {author} {\bibfnamefont {M.}~\bibnamefont
			{{Hanada}}}, \ and\ \bibinfo {author} {\bibfnamefont {M.}~\bibnamefont
			{{Tezuka}}},\ }\bibfield  {title} {\enquote {\bibinfo {title} {{Creating and
					probing the Sachdev-Ye-Kitaev model with ultracold gases: Towards
					experimental studies of quantum gravity}},}\ }\href@noop {} {\bibfield
		{journal} {\bibinfo  {journal} {Prog. Theor. Exp. Phys.}\ }\textbf {\bibinfo
			{volume} {2017}},\ \bibinfo {pages} {083I01} (\bibinfo {year}
		{2017})}\BibitemShut {NoStop}%
	\bibitem [{\citenamefont {{G{\"a}rttner}}\ \emph {et~al.}(2017)\citenamefont
		{{G{\"a}rttner}}, \citenamefont {{Bohnet}}, \citenamefont {{Safavi-Naini}},
		\citenamefont {{Wall}}, \citenamefont {{Bollinger}},\ and\ \citenamefont
		{{Rey}}}]{GarttnerRey:ionOTOC}%
	\BibitemOpen
	\bibfield  {author} {\bibinfo {author} {\bibfnamefont {Martin}\ \bibnamefont
			{{G{\"a}rttner}}}, \bibinfo {author} {\bibfnamefont {Justin~G.}\ \bibnamefont
			{{Bohnet}}}, \bibinfo {author} {\bibfnamefont {Arghavan}\ \bibnamefont
			{{Safavi-Naini}}}, \bibinfo {author} {\bibfnamefont {Michael~L.}\
			\bibnamefont {{Wall}}}, \bibinfo {author} {\bibfnamefont {John~J.}\
			\bibnamefont {{Bollinger}}}, \ and\ \bibinfo {author} {\bibfnamefont
			{Ana~Maria}\ \bibnamefont {{Rey}}},\ }\bibfield  {title} {\enquote {\bibinfo
			{title} {{Measuring out-of-time-order correlations and multiple quantum
					spectra in a trapped-ion quantum magnet}},}\ }\href@noop {} {\bibfield
		{journal} {\bibinfo  {journal} {Nature Physics}\ }\textbf {\bibinfo {volume}
			{13}},\ \bibinfo {pages} {781--786} (\bibinfo {year} {2017})}\BibitemShut
	{NoStop}%
	\bibitem [{\citenamefont {Tsuji}\ \emph {et~al.}(2017)\citenamefont {Tsuji},
		\citenamefont {Werner},\ and\ \citenamefont {Ueda}}]{Tsuji:measurement}%
	\BibitemOpen
	\bibfield  {author} {\bibinfo {author} {\bibfnamefont {Naoto}\ \bibnamefont
			{Tsuji}}, \bibinfo {author} {\bibfnamefont {Philipp}\ \bibnamefont {Werner}},
		\ and\ \bibinfo {author} {\bibfnamefont {Masahito}\ \bibnamefont {Ueda}},\
	}\bibfield  {title} {\enquote {\bibinfo {title} {Exact out-of-time-ordered
				correlation functions for an interacting lattice fermion model},}\
	}\href@noop {} {\bibfield  {journal} {\bibinfo  {journal} {Phys. Rev. A}\
		}\textbf {\bibinfo {volume} {95}},\ \bibinfo {pages} {011601} (\bibinfo
		{year} {2017})}\BibitemShut {NoStop}%
	\bibitem [{\citenamefont {Li}\ \emph {et~al.}(2017)\citenamefont {Li},
		\citenamefont {Fan}, \citenamefont {Wang}, \citenamefont {Ye}, \citenamefont
		{Zeng}, \citenamefont {Zhai}, \citenamefont {Peng},\ and\ \citenamefont
		{Du}}]{Li:NMRmeas}%
	\BibitemOpen
	\bibfield  {author} {\bibinfo {author} {\bibfnamefont {Jun}\ \bibnamefont
			{Li}}, \bibinfo {author} {\bibfnamefont {Ruihua}\ \bibnamefont {Fan}},
		\bibinfo {author} {\bibfnamefont {Hengyan}\ \bibnamefont {Wang}}, \bibinfo
		{author} {\bibfnamefont {Bingtian}\ \bibnamefont {Ye}}, \bibinfo {author}
		{\bibfnamefont {Bei}\ \bibnamefont {Zeng}}, \bibinfo {author} {\bibfnamefont
			{Hui}\ \bibnamefont {Zhai}}, \bibinfo {author} {\bibfnamefont {Xinhua}\
			\bibnamefont {Peng}}, \ and\ \bibinfo {author} {\bibfnamefont {Jiangfeng}\
			\bibnamefont {Du}},\ }\bibfield  {title} {\enquote {\bibinfo {title}
			{Measuring out-of-time-order correlators on a nuclear magnetic resonance
				quantum simulator},}\ }\href@noop {} {\bibfield  {journal} {\bibinfo
			{journal} {Phys. Rev. X}\ }\textbf {\bibinfo {volume} {7}},\ \bibinfo {pages}
		{031011} (\bibinfo {year} {2017})}\BibitemShut {NoStop}%
	\bibitem [{\citenamefont {Yunger~Halpern}\ \emph {et~al.}(2018)\citenamefont
		{Yunger~Halpern}, \citenamefont {Swingle},\ and\ \citenamefont
		{Dressel}}]{YungerHalpern:kindaReview}%
	\BibitemOpen
	\bibfield  {author} {\bibinfo {author} {\bibfnamefont {Nicole}\ \bibnamefont
			{Yunger~Halpern}}, \bibinfo {author} {\bibfnamefont {Brian}\ \bibnamefont
			{Swingle}}, \ and\ \bibinfo {author} {\bibfnamefont {Justin}\ \bibnamefont
			{Dressel}},\ }\bibfield  {title} {\enquote {\bibinfo {title}
			{Quasiprobability behind the out-of-time-ordered correlator},}\ }\href
	{\doibase 10.1103/PhysRevA.97.042105} {\bibfield  {journal} {\bibinfo
			{journal} {Phys. Rev. A}\ }\textbf {\bibinfo {volume} {97}},\ \bibinfo
		{pages} {042105} (\bibinfo {year} {2018})}\BibitemShut {NoStop}%
	\bibitem [{\citenamefont {Yunger~Halpern}(2017)}]{YungerHalpern:Jarzinsky}%
	\BibitemOpen
	\bibfield  {author} {\bibinfo {author} {\bibfnamefont {Nicole}\ \bibnamefont
			{Yunger~Halpern}},\ }\bibfield  {title} {\enquote {\bibinfo {title}
			{Jarzynski-like equality for the out-of-time-ordered correlator},}\
	}\href@noop {} {\bibfield  {journal} {\bibinfo  {journal} {Phys. Rev. A}\
		}\textbf {\bibinfo {volume} {95}},\ \bibinfo {pages} {012120} (\bibinfo
		{year} {2017})}\BibitemShut {NoStop}%
	\bibitem [{\citenamefont {Garc\'{\i}a-Garc\'{\i}a}\ \emph
		{et~al.}(2018)\citenamefont {Garc\'{\i}a-Garc\'{\i}a}, \citenamefont
		{Loureiro}, \citenamefont {Romero-Berm\'udez},\ and\ \citenamefont
		{Tezuka}}]{Garcia:SYKtransition}%
	\BibitemOpen
	\bibfield  {author} {\bibinfo {author} {\bibfnamefont {Antonio~M.}\
			\bibnamefont {Garc\'{\i}a-Garc\'{\i}a}}, \bibinfo {author} {\bibfnamefont
			{Bruno}\ \bibnamefont {Loureiro}}, \bibinfo {author} {\bibfnamefont
			{Aurelio}\ \bibnamefont {Romero-Berm\'udez}}, \ and\ \bibinfo {author}
		{\bibfnamefont {Masaki}\ \bibnamefont {Tezuka}},\ }\bibfield  {title}
	{\enquote {\bibinfo {title} {Chaotic-integrable transition in the
				sachdev-ye-kitaev model},}\ }\href {\doibase 10.1103/PhysRevLett.120.241603}
	{\bibfield  {journal} {\bibinfo  {journal} {Phys. Rev. Lett.}\ }\textbf
		{\bibinfo {volume} {120}},\ \bibinfo {pages} {241603} (\bibinfo {year}
		{2018})}\BibitemShut {NoStop}%
	\bibitem [{\citenamefont {{Basko}}\ \emph {et~al.}(2006)\citenamefont
		{{Basko}}, \citenamefont {{Aleiner}},\ and\ \citenamefont
		{{Altshuler}}}]{BAA}%
	\BibitemOpen
	\bibfield  {author} {\bibinfo {author} {\bibfnamefont {D.~M.}\ \bibnamefont
			{{Basko}}}, \bibinfo {author} {\bibfnamefont {I.~L.}\ \bibnamefont
			{{Aleiner}}}, \ and\ \bibinfo {author} {\bibfnamefont {B.~L.}\ \bibnamefont
			{{Altshuler}}},\ }\bibfield  {title} {\enquote {\bibinfo {title} {{Metal
					insulator transition in a weakly interacting many-electron system with
					localized single-particle states}},}\ }\href@noop {} {\bibfield  {journal}
		{\bibinfo  {journal} {Ann. Phys.}\ }\textbf {\bibinfo {volume} {321}},\
		\bibinfo {pages} {1126} (\bibinfo {year} {2006})}\BibitemShut {NoStop}%
	\bibitem [{Note1()}]{Note1}%
	\BibitemOpen
	\bibinfo {note} {See Ref.~\protect \rev@citealpnum {NandkishoreHuse:review}
		for a review}\BibitemShut {NoStop}%
	\bibitem [{\citenamefont {Abrikosov}\ \emph {et~al.}(1975)\citenamefont
		{Abrikosov}, \citenamefont {Gorkov},\ and\ \citenamefont
		{Dzyaloshinski}}]{AGD}%
	\BibitemOpen
	\bibfield  {author} {\bibinfo {author} {\bibfnamefont {A.~A.}\ \bibnamefont
			{Abrikosov}}, \bibinfo {author} {\bibfnamefont {L.~P.}\ \bibnamefont
			{Gorkov}}, \ and\ \bibinfo {author} {\bibfnamefont {I.~E.}\ \bibnamefont
			{Dzyaloshinski}},\ }\href@noop {} {\emph {\bibinfo {title} {Methods of
				Quantum Field Theory in Statistical Physics}}}\ (\bibinfo  {publisher}
	{Dover, New York},\ \bibinfo {year} {1975})\BibitemShut {NoStop}%
	\bibitem [{\citenamefont {Kamenev}(2011)}]{Kamenev:book}%
	\BibitemOpen
	\bibfield  {author} {\bibinfo {author} {\bibfnamefont {A.}~\bibnamefont
			{Kamenev}},\ }\href@noop {} {\emph {\bibinfo {title} {Field Theory of
				Non-Equilibrium Systems}}}\ (\bibinfo  {publisher} {Cambridge Univ. Press},\
	\bibinfo {address} {Cambridge},\ \bibinfo {year} {2011})\BibitemShut
	{NoStop}%
	\bibitem [{\citenamefont {Syzranov}\ \emph {et~al.}(2018)\citenamefont
		{Syzranov}, \citenamefont {Gorshkov},\ and\ \citenamefont
		{Galitski}}]{Syzranov:OTOCdot}%
	\BibitemOpen
	\bibfield  {author} {\bibinfo {author} {\bibfnamefont {S.~V.}\ \bibnamefont
			{Syzranov}}, \bibinfo {author} {\bibfnamefont {A.~V.}\ \bibnamefont
			{Gorshkov}}, \ and\ \bibinfo {author} {\bibfnamefont {V.}~\bibnamefont
			{Galitski}},\ }\bibfield  {title} {\enquote {\bibinfo {title}
			{Out-of-time-order correlators in finite open systems},}\ }\href {\doibase
		10.1103/PhysRevB.97.161114} {\bibfield  {journal} {\bibinfo  {journal} {Phys.
				Rev. B}\ }\textbf {\bibinfo {volume} {97}},\ \bibinfo {pages} {161114}
		(\bibinfo {year} {2018})}\BibitemShut {NoStop}%
	\bibitem [{\citenamefont {Sachdev}\ and\ \citenamefont {Ye}(1993)}]{SachdevYe}%
	\BibitemOpen
	\bibfield  {author} {\bibinfo {author} {\bibfnamefont {Subir}\ \bibnamefont
			{Sachdev}}\ and\ \bibinfo {author} {\bibfnamefont {Jinwu}\ \bibnamefont
			{Ye}},\ }\bibfield  {title} {\enquote {\bibinfo {title} {Gapless spin-fluid
				ground state in a random quantum heisenberg magnet},}\ }\href@noop {}
	{\bibfield  {journal} {\bibinfo  {journal} {Phys. Rev. Lett.}\ }\textbf
		{\bibinfo {volume} {70}},\ \bibinfo {pages} {3339--3342} (\bibinfo {year}
		{1993})}\BibitemShut {NoStop}%
	\bibitem [{\citenamefont {{Nandkishore}}\ and\ \citenamefont
		{{Huse}}(2015)}]{NandkishoreHuse:review}%
	\BibitemOpen
	\bibfield  {author} {\bibinfo {author} {\bibfnamefont {R.}~\bibnamefont
			{{Nandkishore}}}\ and\ \bibinfo {author} {\bibfnamefont {D.~A.}\ \bibnamefont
			{{Huse}}},\ }\bibfield  {title} {\enquote {\bibinfo {title} {{Many-Body
					Localization and Thermalization in Quantum Statistical Mechanics}},}\
	}\href@noop {} {\bibfield  {journal} {\bibinfo  {journal} {Annual Review of
				Condensed Matter Physics}\ }\textbf {\bibinfo {volume} {6}},\ \bibinfo
		{pages} {15--38} (\bibinfo {year} {2015})}\BibitemShut {NoStop}%
	\bibitem [{\citenamefont {Klug}\ and\ \citenamefont
		{Syzranov}(2019)}]{Syzranov:OTOCdetail}%
	\BibitemOpen
	\bibfield  {author} {\bibinfo {author} {\bibfnamefont {M.}~\bibnamefont
			{Klug}}\ and\ \bibinfo {author} {\bibfnamefont {S.~V.}\ \bibnamefont
			{Syzranov}},\ }\href@noop {} {} (\bibinfo {year} {2019}),\ \bibinfo {note}
	{coming soon}\BibitemShut {NoStop}%
\end{thebibliography}
\end{document}